\documentclass[letterpaper]{article} %
\usepackage[preprint]{aaai2027}  %
\usepackage[hyphens]{url}  %
\usepackage{graphicx} %
\usepackage{natbib}  %
\usepackage{caption} %
\usepackage{booktabs}
\usepackage{amsmath,amssymb}
\usepackage{multirow}

\title{Unread or Unenforced? Separating Representation from\\
Enforcement Failure in Content Guards}

\author{
    Haoyu Zhang\textsuperscript{\rm 1},
    Yi Feng\textsuperscript{\rm 2},
    Shibo Zheng\textsuperscript{\rm 1},
    Hanwen Liu\textsuperscript{\rm 3},
    Haowen Xu\textsuperscript{\rm 1},
    Xiao Luo\textsuperscript{\rm 1},
    Zhuoxi Wang\textsuperscript{\rm 1},
    Mohammad Zandsalimy\textsuperscript{\rm 1},
    Shanu Sushmita\textsuperscript{\rm 2}
}
\affiliations{
    \textsuperscript{\rm 1}Individual Contributor\\
    \textsuperscript{\rm 2}Northeastern University\\
    \textsuperscript{\rm 3}University of Southern California\\
    zhang.haoyu6@northeastern.edu
}

\begin{document}
\maketitle

\begin{abstract}
When an encoded attack passes a content guard, the guard either never
represented the payload's harmful content or represented it and failed to act.
End-to-end attack success rate reports one number for both, yet the two have
opposite remedies: one is a representational limit that more verdict-level safety training
cannot reach, the other is a decision rule that it can. We separate them by
reading a guard's own residual stream, using a content probe fitted on plaintext
and transferred without refitting to the encoded condition, alongside the
verdict logits from the same pass. The measurement is deterministic: one forward
pass per prompt, no sampling, no generation, no judge model.

Licensing that read honestly is most of the problem and is our main
contribution. A conventional permutation test admits the decode measurement on
most of a 19-condition encoding ladder for each of two open guards. A
length-matched null and a floor calibrated on conditions the guard's base model
decodes at rate zero reduce it to four conditions each; holding out the items
the probe was fitted on removes one more. The discarded cells are not marginal
ones. The largest result in our first analysis was a
guard apparently representing a cipher while blocking none of it; that
encoding is one its base model decodes at rate zero, and three further cells
share the signature of high apparent decode with a block rate of exactly zero. A
further screen constrains the block axis, which the decode screens leave
untouched, by running plaintext content inside each condition's own wrapper. It
removes the largest cell that survived them. A length bound on the same axis,
measured in characters and in each guard's own tokens, leaves every reported
cell standing.

What remains is a policy failure that survives an item-level holdout on one
encoding (text interleaved with invisible zero-width characters) for both guards, at 8 and 7 per 100 prompts, against 17 and 23 when the
probe is allowed to have seen the prompt it is scoring. It is also confined to one family of surface encodings: where
an encoding leaves content linearly recoverable we can separate the two
failures, and on genuine ciphers we report the cells as unmeasured rather than
as evidence that nothing was decoded. Across
every guard and condition pair except two, both unmeasured on the decode axis, blocked without decoding is
near zero, so we find little evidence for a pure encoding-format detector under the conditions we test. That cell is the
one read we do not repeat under the holdout, and we report
it as such.
\end{abstract}

\section{Introduction}

Content guards are the deployed answer to unsafe requests. A separate
classifier (a member of the Llama Guard family
\citep{inan2023llamaguardllmbasedinputoutput}, of which we run the version-3 8B
checkpoint \citep{grattafiori2024llama3}, or WildGuard
\citep{NEURIPS2024_0f69b4b9}) reads a prompt before or alongside the generating
model and returns a safety verdict. The field evaluates such guards end to end: a corpus
of attacks is passed through, and the fraction that survives is reported as an
attack success rate or, equivalently, as one minus a block rate.

That single number is a composition of two distinct failures, and it does not
say which occurred. When an encoded attack (a prompt whose harmful payload has
been transformed by a cipher, a homoglyph substitution, or an invisible-character
scheme) passes a guard, exactly one of two things happened. Either the guard
never represented the payload's content at all, in which case its verdict was
made on a string it could not read; or it did represent the content and returned
\emph{safe} anyway. We call these the \emph{capability} failure and the
\emph{policy} failure.

The distinction is not academic, because the two have opposite remedies. A
capability failure is a representational limit: the guard's forward pass never
recovers the harmful semantics, and no amount of additional preference data over
verdicts will fix it, because there is nothing in the residual stream for a
better decision rule to act on. A policy failure is the reverse: the content is
present and legible, and what fails is the mapping from content to verdict, which
is precisely what more or better safety training addresses. An evaluation that
reports only the composed rate tells a defender that their guard is leaking
without telling them which of two disjoint engineering programmes to fund.

We separate the two by reading the guard's own activations. A linear content
probe fitted on plaintext contrasts is evaluated, without refitting, on the
guard's forward pass over the encoded prompt; the guard's verdict is read from
the logits at its own verdict position in the same pass. Crossing the two yields
four cells, of which \emph{decoded but not blocked} is the policy failure and
\emph{never decoded} is the capability failure. We use those names throughout for
the measured cells. Establishing that the second is an enforcement failure in the
causal sense, that the verdict computation could have used content it did not,
requires intervening on the recovered direction, which we do not do here; what we
report is that the content is linearly recoverable and the guard returned safe. The measurement needs no
generation, no sampling and no judge model, so it costs one forward pass per
prompt.

Doing this honestly turns out to be most of the work, and it is the part we
consider the contribution. A content probe read on encoded text is a
notoriously permissive instrument: it will fire on surface properties that
correlate with harmfulness in the evaluation corpus but carry no decoded content
whatsoever. We show that under free permutation licensing (the natural significance test), 17 of 19 encoding conditions on one guard and 12 of 19 on the
other appear measurable, and that under a length-matched null the honest answers
are 6 and 7. Without that control this paper would have reported that both guards
decode most of a cipher ladder and let the majority through, a headline built on
a probe reading character length. A second control, calibrated against encoding
conditions the guard's base model decodes at rate zero, removes further conditions,
among them the largest cell in our first map.

The result that survives is narrower than the one we set out to find and, we
think, more useful. Where decode is measurable, the policy failure survives the
item-level holdout on one encoding, text interleaved with invisible zero-width
characters, for both guards. The cipher band is unmeasured: neither guard
blocks more than 2 per 100 there, and we cannot establish that either decoded anything. And
\emph{blocked without decoding} (the cell that would indicate a guard reacting
to the mere appearance of encoding rather than to content) is empirically near
zero across every guard--condition pair except two, both unmeasured on the
decode axis. We find little evidence for
a pure encoding-format detector under the conditions we test, though the reading
is contingent on a linear probe at one position, on a licensing procedure that
leaves most cipher conditions unmeasured, and on a decode read that is the one
cell here not taken under the item-level holdout, for the reason given in
Section~\ref{sec:format}.

\section{Scope}
\label{sec:scope}

We state the paper's limits before its results, because several of them shaped
what we chose to measure.

\paragraph{Two guards, one checkpoint each.} We study Llama Guard~3~(8B) and
WildGuard~(7B). Two is the minimum number at which a dissociation between guards
is a dissociation rather than an anecdote, and we do report such dissociations.
It is not a number at which either guard's individual behaviour becomes a
general claim about content guards, and we do not make one.

\paragraph{The decode axis is measurable only where the encoding is
surface-level.} Our decode measurement is a linear probe, and a linear probe
recovers content only if the encoded representation carries it in a
linearly accessible form. On the conditions where this holds (zero-width
characters, homoglyph substitution, word-order permutation, and combining marks
on one guard), the decode axis is measurable. On genuine ciphers it is not: the probe has no
signal above the length baseline, and we therefore report those conditions as
\emph{unmeasured} rather than as \emph{not decoded}. This is a real restriction
on the paper's reach and we mark it everywhere it applies. It is not a null result about cipher decoding; it is the absence of an
instrument.  Before the item-level holdout the measurable band is four conditions per guard
(the holdout then removes one of Llama Guard's, and the block-axis screen one of
WildGuard's), and the two
conditions that replicate across both guards, homoglyph substitution and zero-width characters, belong to the same family. Our separation of the two failure modes
is therefore demonstrated on surface Unicode substitution, not on encoded
prompts in general.

\paragraph{The measurement is correlational.} A probe that reads harmful content
from a guard's residual stream establishes that the information is linearly
present at that layer and position. It does not establish that the guard's
verdict computation uses it. Causal tests, ablating the direction and observing
the verdict, are the natural next step and are outside this paper.

\paragraph{We make no claim about target-model behaviour.} This paper measures
what happens inside the \emph{defence}. Whether a generating model subjected to
the same encoded prompts refuses, complies, or fails to decode them is a
separate question about a different system, and nothing here should be read as
evidence about it.

\paragraph{One corpus, English.} All conditions are applied to a single pair of
harmful and benign prompt sets. The encodings are applied to the payload only,
leaving the surrounding chat template intact, which is the deployment-realistic
case but also means our conditions share whatever idiosyncrasies that corpus
has.

\section{Related work}

\paragraph{Guards, and how they are evaluated.} Dedicated safety classifiers
\citep{inan2023llamaguardllmbasedinputoutput,grattafiori2024llama3,%
NEURIPS2024_0f69b4b9,sharma2025constitutional} are evaluated end to end: a
corpus of attacks is passed through and the surviving fraction is reported
\citep{jalan2026survey}. That protocol is what our decomposition responds to; it
is not itself in dispute, and for the many questions that concern deployment
risk it is the right number. \citet{xin-etal-2026-jailbreaking} run it across
the full inference pipeline, input and output filtering included, and conclude
that nearly every evaluated jailbreak is caught by at least one filter. We take
that result at face value and note only what it cannot separate: a pipeline that
blocks well may be blocking because the guard recovered the content, or because
the guard reacts to the shape of an attack it never read. Those two have the
same block rate and different failure modes under distribution shift.

\paragraph{Encoded prompts against guards.} That encodings defeat safety
behaviour is established, and we claim none of it. \citet{wei2023jailbroken}
attribute such failures to mismatched generalization between a model's
capabilities and its safety training, and DecipherGuard
\citep{yang2025decipherguardunderstandingdecipheringjailbreak} demonstrates the
failure on guards specifically and repairs much of it by decoding the prompt
before classification. Our question begins where that result ends. A guard
repaired by an external decoding step tells us decoding was missing somewhere in
the pipeline; it does not tell us whether the guard had already recovered the
payload internally and failed to act on it.

One of our two cells has a stronger prior than that. \citet{%
fairoze2026bypassingpromptguardsproduction} start from an impossibility result,
that no filter running materially faster than the model it protects can
universally separate adversarial from benign prompts, and construct prompts that
are indecipherable to a bounded filter while remaining tractable to the target
model, defeating four production systems and fourteen open-weight guards. That
is our \emph{never decoded} mechanism built on purpose and with a
lower bound behind it, and we claim no novelty for the mechanism itself. What we
add is the complementary question: under ordinary surface encodings, which
nobody designed to be undecodable, how often does a guard fail because it never
recovered the payload, and how often does it recover the payload and let it
through anyway? Their construction guarantees the first case; ours has to
measure which case occurred.

\paragraph{Inside a guard.} SIREN \citep{jiao2026siren} builds a harmfulness
detector out of internal representations and is the closest prior work to ours in
method, but it constructs a better guard rather than auditing an existing one,
and evaluates no encoded inputs. \citet{liu2026cognition} likewise build rather
than audit, fusing layer-wise probes from several depths; their contribution is
that a single depth is not enough, while our decode read is taken at one selected
layer and position, which we adopt because a selected maximum is what our
permutation null and control floor are built to correct for. Their reported
separability is measured with a fitted probe rather than a transferred one, and
our item-level holdout is the control that would tell the two apart. Gamma-Guard
\citep{lv-etal-2025-gamma} traces guard failure to representation shifts
qualitatively, with no encoding conditions and no
representation-versus-enforcement distinction. Two recent results establish that
a guard's verdict can track a surface cue rather than content:
\citet{tasawong-etal-2025-shortcut} find prompt-side keyword bias correlated with
training labels, and \citet{feng2026refusalcueshortcut} find a response-side
refusal cue that flips a verdict without altering the harmful content, in a
dataset our own guards are trained on. Both are reasons a block rate needs a
control holding content fixed, which is what our wrapper arm provides.

\paragraph{Reading content out of activations.} Our decode measurement is a
linear read of the residual stream, in the line of the difference-in-means directions
of \citet{arditi2024refusal}, though our estimator is a logistic probe
(supplementary material). \citet{zhao2025llmsencode} show that a model's
internal representation of harmfulness is separable from its refusal decision,
and localise the two at the instruction-final and post-instruction tokens
respectively. Those are the two positions we capture, and that is the published
interpretation of what each carries. \citet{youstra2025cifr} train probe monitors on internal
activations to detect cipher-encoded content in fine-tuning data; that is the
closest existing pairing of encodings with internal probes, on a defence problem
one layer removed from ours.

\section{Method}

\subsection{Guards and conditions}

Both guards are open-weight classifiers with fixed prompt formats. We render
each prompt through the guard's own chat template and tokenise without adding
special tokens, because both templates emit their own. Each harmful and benign
prompt is presented in plaintext and under each encoding condition; the encoding
is applied to the payload, never to the template. The corpus is
JailbreakBench's 100 harmful behaviours together with its topic-matched benign
counterparts \citep{NEURIPS2024_63092d79}; the matching matters, because an
unmatched negative class would let a content probe separate on topic and be
read as content recovery. Each condition is run on all one hundred harmful and
all one hundred benign prompts, so every cell count we report is out of one
hundred.

\subsection{Measurement 1: does the guard represent the content?}

We fit a linear probe on the guard's residual stream to separate harmful from
benign \emph{plaintext} prompts, and evaluate it, without refitting, on the
guard's forward pass over the \emph{encoded} prompt. Refitting on encoded
activations would be circular: a probe trained to separate encoded-harmful from
encoded-benign can succeed on any feature that distinguishes the two corpora,
decoded content included but not required \citep{belinkov2022probing}. Transfer from a plaintext-fitted
probe is the weaker and therefore the meaningful direction.

The probe is read at two positions (the last token of the payload and the last
token of the rendered prompt) and swept over every layer; the reported
statistic is the transfer AUROC at the best licensed layer--position cell.
Selecting that cell by its own score would inflate the statistic, so the
selection sits inside the test: the maximum transfer AUROC over the grid is
compared against a null of maxima obtained by permuting the \emph{evaluation}
labels with the fitted direction held fixed, which makes licensing one test
rather than one per cell. Permuting the training labels instead is the natural
first choice and is wrong here: a logistic fit on shuffled labels is
near-constant, so its transfer AUROC collapses onto $0.5$ with almost no
variance and the resulting null is far too tight. A per-prompt read is obtained
by thresholding against the \emph{same-condition} benign score distribution, so
that the encoding and its template wrapper are common-mode and cannot by
themselves decide a label. The supplementary material states the estimator, the
grid, the split sizes, the permutation count and the thresholds in full.

\subsection{Measurement 2: what verdict did the guard return?}

Both guards answer in a fixed format, so the verdict is read directly from the
logits at the position where the verdict token falls: the guard blocks when the
unsafe label outscores the safe one. This requires one forward pass, no
sampling, and no external judge, which makes the whole map free of the
LLM-as-judge failure modes that afflict end-to-end evaluation, and free in money.

Two details are load-bearing and were each found by a failure. The verdict
position is not token zero: one guard emits two newlines before its verdict, and
reading at the naive position scored the labels where they held essentially none
of the probability mass, reporting a block rate of zero for a guard that blocks
almost everything. We therefore report, for every run, the mean probability mass
sitting on the two label tokens at the read position, and treat a value away
from one as a format failure rather than a finding. Second, the whitespace
belongs on the label rather than on the prefix, because the tokeniser encodes the
boundary into the token itself.

\subsection{The four cells}

Crossing the two measurements gives \emph{blocked on content}, \emph{decoded but
not blocked}, \emph{blocked without decoding}, and \emph{never decoded}. When the
decode measurement is not licensed for a condition, the cell is
\emph{unmeasured}: the decode reading is a three-valued quantity, and collapsing
its unmeasured state onto ``did not decode'' would assert a capability failure
that was never observed. This is the single most consequential design choice in
the instrument.

\subsection{Controls}

The controls are as much the contribution as the map, because several of them
removed something we had already believed.

\paragraph{A length-matched null.} Harmful and benign prompts differ in length in
our corpus, as they do in the standard ones; every encoding is monotone in
length, so the difference survives into every condition, and a probe can achieve
a substantial transfer AUROC by reading length alone. We therefore license a
condition not against a free permutation of labels but against a permutation
stratified by length, and we verify that the licensing decision is stable across
5, 10 and 20 strata.

\paragraph{A control floor.} Significance is not sufficiency: at our sample size
a condition can clear a permutation null while reading barely above what the
same probe reads on conditions carrying no decoded content at all. In the spirit
of the control tasks of \citet{hewitt2019designing}, applied to a transferred
probe, we therefore
derive a floor from the conditions the guard's \emph{base model} decodes at
rate 0.00 (the guard itself cannot be asked to restate a payload, because
its template hard-wires the classification task) and require a condition to
clear that floor by a margin, or be reported as unmeasured.

\paragraph{An item-level holdout.} Transferring a probe without refitting holds
out the \emph{condition} but not the \emph{items}: the encoded corpus is the
same prompts wearing a transformation, so a probe told that plaintext item $i$
is harmful can score encoded item $i$ by recognising the item rather than by
recovering its content. At our sample size the fit interpolates, which makes
that channel available. We therefore split the corpus by item as well as by
condition (fit on the plaintext half, evaluate on the encoded versions of the
held-out half, average over many random splits) and report the held-out
statistic. The size of the leak is measured directly by scoring the same probe
on the encoded versions of the items it was fitted on, which isolates item
memory from the smaller training set. The same screens are re-derived under the
holdout, floors included, so that a condition and the floor it must clear are
computed the same way.

\paragraph{A benign arm.} Every block rate is a rate with no denominator unless
the same guard is run on benign content in the same encoding. Without it, a guard
that flags anything wearing an unusual format produces a perfect block rate that
means nothing.

\paragraph{A wrapper arm.} An encoded condition changes two things at once: the
payload's characters, and the template announcing that an encoding is present.
The benign arm above holds the encoding fixed and varies harm; this holds the
content fixed as plaintext and varies only the wrapper, the format-decorrelation
design of \citet{devbunova2026formatsensitivity}, so that a guard
responding to the \emph{request to decode} can be distinguished from one
responding to what it recovered. We take the floor from the benign arm of that
condition (safe content, wrapped, blocked anyway) rather than the harmful one,
since blocking wrapped harmful plaintext is correct behaviour and would penalise
a working guard. This control is not shared with the decode screens and is the
only block-axis control that holds content fixed.

\paragraph{A plaintext baseline.} Both arms above are rates within an encoded or
wrapped condition, and comparing them across guards requires knowing what each
guard does with the bare corpus. We run every guard on unencoded, unwrapped
harmful and benign prompts; the two guards differ by 22 points in benign block
rate before any encoding is applied, so an encoded rate quoted without this
denominator attributes a checkpoint difference to the attack.

\paragraph{A length bound on the block axis.} The length-matched null above
licenses \emph{probes}, and a probe AUROC is the only quantity it reaches. Block
rates are not probe readings, and none of the arms above constrains them for
length: the benign arm holds the encoding fixed and varies harm, the wrapper arm
holds the content fixed and varies the template, and both leave the two arms'
lengths free to differ exactly as the bare corpora do. We therefore bound the
harm gap that length alone can produce. Fix the blocking budget to the number of
prompts the real guard blocked across both arms, and give that budget to the best
monotone length-only rule, which blocks the longest or the shortest items of the
pooled set, whichever separates more. Its harm gap is the most a guard reacting
only to length could show at this guard's own operating rate. We compute it in
characters and again in the guard's own tokens, taking the larger as binding,
because the two come apart on exactly the conditions where an encoding inflates
sequence length without changing the character count. All six reported cells
exceed their bound, by $0.32$, $0.32$ and $0.14$ for Llama~Guard on
\texttt{homoglyph}, \texttt{zero\_width} and \texttt{reverse\_words}, and by
$0.26$, $0.26$ and $0.10$ for WildGuard. The token bound is the binding one on
three of the six, and on WildGuard \texttt{reverse\_words} it exceeds the
character bound by $0.12$, which is more than the margin that survives it; a
control measured only in characters would have reported that cell as
comfortable. The bound's construction, its restriction to monotone rules, and its
insensitivity to the blocking budget are given in the supplementary material.

\paragraph{An operating-point sweep.} The per-prompt decode read is thresholded,
so every cell count is a function of that threshold. We report the sweep rather
than a single point, and we show which counts are invariant under it and which
are not.

\section{Results}

\subsection{Most of the apparent map is the controls' doing}

\begin{table}[t]
\centering
\setlength{\tabcolsep}{3pt}
\begin{tabular}{lcccc}
\toprule
& \multicolumn{2}{c}{Llama Guard 3} & \multicolumn{2}{c}{WildGuard} \\
\cmidrule(lr){2-3}\cmidrule(lr){4-5}
Screen & admits & undec. & admits & undec. \\
\midrule
Free permutation null & 17 & $\geq 10$ & 12 & $\geq 7$ \\
Length-matched null & 6 & 1 & 7 & 3 \\
\quad + control floor & 4 & 0$^*$ & 4 & 0$^*$ \\
\quad + item-level holdout & \textbf{3} & 0$^*$ & \textbf{4} & 0$^*$ \\
\bottomrule
\end{tabular}
\caption{Encoding conditions (of 19) on which the decode measurement is
licensed, by screen. The first row is what a conventional significance test
admits; the last is what survives. \emph{undec.} counts the admitted conditions
that the guard's base model cannot decode at all (ability $0.00$), on which,
assuming the guard inherits that inability, nothing was decoded and a licence is therefore a false positive by a
criterion that never consults the probe. The free-null entries are lower bounds
from the marginal counts, since only 7 and 5 of the 19 conditions are decodable
at all. $^*$Zero under the control floor is partly by construction: the floor's
control set is those same undecodable conditions.}
\label{tab:attrition}
\end{table}

Table~\ref{tab:attrition} is the paper's first result, and it is a negative one
about method. A free permutation test, the natural way to license a probe,
admits 17 of 19 conditions on one guard and 12 of 19 on the other. Reported at
that stage, this study would have concluded that both guards represent the
content of most of a cipher ladder while blocking almost none of it. Under a
length-matched null the count falls to 6 and 7, under the control floor to 4 and
4, and under an item-level holdout to 3 and 4.

\paragraph{Removing conditions is not the same as improving validity.} A screen
that only ever rejects can be tightening a measurement or merely shrinking it,
and the counts alone do not distinguish those. One criterion here is independent
of every screen in this paper. If a guard's base model cannot decode a condition
at all, then, assuming the guard inherits that inability, nothing was decoded and
a licence granted there is a false positive, and
the decoding ability is measured by generation rather than by any probe, null or
floor. Twelve of the nineteen conditions are undecodable by Llama-3.1-8B and
fourteen by Mistral-7B-Instruct-v0.3, which leaves only 7 and 5 that could
carry a true positive at all. A free permutation test therefore buys at least 10
and at least 7 false positives before any of our screens run; the length-matched
null cuts that to 1 and 3.

\paragraph{One row is not an independent gain, and we say so.} The control floor
drives the false positives to zero, but its control set \emph{is} the
undecodable conditions, so that zero is the criterion applied rather than a test
passed, and reporting it as a score would be circular. What the floor adds over
the ability criterion is coverage of conditions where ability is partial rather
than absent, which is where a decodability threshold gives no answer. The
item-level holdout is the screen with the cleanest claim here: it consults
neither ability nor the control set, it is the last screen applied, and it still
removes a condition that had passed everything before it.

\subsection{Significance is not sufficiency}

The clearest instance is a set of conditions that pass the free permutation test
comfortably and are nonetheless artefacts. On Llama Guard, one Caesar-shift
condition passes the length-matched null at the smallest p-value 200 draws can produce, with an
apparent \emph{decoded but not blocked} rate of 0.77, the largest cell in our
first analysis and the one the study was initially built around. On WildGuard, three conditions that also pass the
length-matched null, though not at that minimum, read 0.76, 0.69
and 0.66. All four sit on encodings the corresponding base model decodes at rate
0.00, so none of them is evidence of a policy failure; we read them as the probe
picking up surface features (on the Caesar-shift cell, a few thousandths above
its own control distribution).

These cells share a signature worth naming, because it will recur in any
uncontrolled version of this measurement: \emph{high apparent decode rate,
block rate exactly zero}. That combination is what a false positive looks like
here, and it is also what the most publishable finding looks like.

Two screens sharing no input, the control floor and the stability of licensing
under the number of strata, recover the same partition on WildGuard and not on
Llama Guard, where the Caesar-shift artefact is stable at every bin count and is
caught only by the floor; bin-stability is therefore the weaker screen and cannot
substitute for it. The supplementary material gives both partitions.

\subsection{The probe was partly reading the item, and the leak is
ability-dependent}
\label{sec:holdout}

The transfer design holds out the condition and not the items, so we measured
what that is worth. Fitting on half the plaintext items and evaluating on the
encoded versions of the other half, averaged over 200 random splits, moves every
reported AUROC down: the largest, 0.985, becomes 0.845.

The size of the leak is not uniform, and its pattern is the reason the control
floor could not have caught it. Scoring the same probe, at the same training-set
size, on items it was fitted on rather than on held-out ones gives a direct
estimate of item memory. On seven surface-encoding cells (homoglyph, zero-width and
full-width on both guards, combining marks on WildGuard), that estimate is 0.147 to 0.238 of AUROC. On the conditions used as
controls, whose base-model restatement rate is zero, it is $-0.011$ to $0.048$. Item memory
requires the transformation to preserve the item's identity in the
representation, which is what decoding it does; where nothing is decoded, there is little to remember.  The floor is estimated exactly where the
confound is weakest and applied exactly where it is strongest, so it moves by
0.003 on one guard and 0.020 on the other while the conditions we would otherwise have reported fall by 0.04 to
0.20. A screen can be sound and still be structurally blind to a confound, and
this one was.

Because the collapse is differential rather than proportional, most of the map
survives it. Six of the seven conditions we would otherwise have reported still
clear their re-derived floor. One does not: on Llama Guard, \texttt{fullwidth}
reads 0.706 against a floor of 0.707. We do not report that as a demotion by a
thousandth, because the spread of that condition's held-out AUROC across splits
runs from 0.63 to 0.79 and straddles the floor. It joins the \emph{unmeasured}
band, and on the other guard a condition clearing by 0.004 on the far side of
the same line is reported the same way, as passing rather than as established.
The comparison that does not depend on a thousandth is with the controls
themselves: under the holdout that condition reads \emph{below} the
Caesar-shift condition of the previous subsection, on which the base model
decodes nothing at all. It is the cell rather than the margin that has to go.

The holdout also supplies a denominator the design had been missing. A
cross-validated probe fitted and evaluated \emph{within} plaintext reaches 0.90
to 0.98 at the same cells, so the held-out encoded reading is degraded rather
than reduced to the floor: on Llama Guard's
\texttt{homoglyph}, the highest-AUROC cell, by 0.09 (0.845 against a plaintext
0.935), and by more on the others. The content survives the
encoding into the guard's residual stream at a measurable cost, which is a
stronger statement than the uncontrolled 0.985 supported, because 0.985 was
partly the probe recognising a prompt it had already been shown.

\subsection{The map that survives}

\begin{table}[t]
\centering
\small
\setlength{\tabcolsep}{1.5pt}
\begin{tabular}{llccc}
\toprule
Guard & Condition & AUROC & D\&$\neg$B & Block \\
\midrule
\multirow{3}{*}{Llama Guard 3}
 & \texttt{homoglyph}       & 0.845 & 3.3 [0--8] & 0.92 \\
 & \texttt{zero\_width}     & 0.767 & 8.1 [2--16] & 0.83 \\
 & \texttt{reverse\_words}$^\dagger$ & 0.758 & 5.0 [0--14] & 0.65 \\
\midrule
\multirow{3}{*}{WildGuard}
 & \texttt{homoglyph}       & 0.789 & 4.1 [0--10] & 0.75 \\
 & \texttt{zero\_width}     & 0.772 & 6.8 [2--14] & 0.71 \\
 & \texttt{reverse\_words}$^\dagger$ & 0.764 & 2.7 [0--8] & 0.73 \\
\bottomrule
\end{tabular}
\caption{Conditions surviving all screens. AUROC is the item-held-out transfer
statistic at the selected layer--position cell, against re-derived floors of
0.707 (Llama Guard 3) and 0.661 (WildGuard). D\&$\neg$B is the rate, per 100
harmful prompts, that the guard represented and did not block, at the tuned
operating point, under the same item-level holdout as the AUROC beside it: the
probe is refitted on half the plaintext items and the per-prompt read is taken on
the held-out half. We report the mean over 200 splits and the 2.5th and 97.5th
percentiles across them, which is a spread across splits rather than a confidence
interval. The cell requires a \emph{positive} decode read, so a probe denied item
memory tends to under-count it and we read each value as a lower bound; two of the four
non-control cells exclude zero. An unsplit probe reads the same six cells at 7,
17, 8, 23, 23 and 9, 1.6 to 5.6 times higher, and its decode term is $0.94$ to
$0.99$ on the four non-control cells, against $0.63$ to $0.79$ held out across
all six. We do not order the conditions by this
count: the split spreads overlap on every pair, and a decode term that far below
one is no longer safely the non-binding term of the conjunction, so an ordering
would mix comprehension with policy.
$^\dagger$\texttt{reverse\_words} leaves the payload's words
lexically intact and is reported as a control, not a finding. Llama Guard's
\texttt{fullwidth} appeared here before the item-level holdout and is now
reported as unmeasured on the decode axis; one further WildGuard condition
passes the decode screens and is excluded by the wrapper screen below.}
\label{tab:map}
\end{table}

The policy failure survives on \texttt{zero\_width} for both guards
(Table~\ref{tab:map}), at 8 and 7 per 100 on a condition the guards block
heavily; on \texttt{homoglyph} the split spread includes zero. It is a narrower claim than the uncontrolled analysis supported, and
narrower again than our own analysis supported one screen ago. The two guards
also dissociate completely on one condition: on \texttt{fullwidth} Llama Guard
blocks 85 of 100 harmful prompts and WildGuard blocks 0, on the same corpus, and
neither guard's decode measurement there survives our screens, so we report the
behaviour and not what either guard represents (supplementary material).

The rest of the ladder is unmeasured on the decode axis. Neither guard blocks
more than 2 per 100 of any genuine cipher condition (the supplementary material tabulates every block rate),
and on none of them can we establish that either guard decoded anything; those
cells are \emph{unmeasured}, not \emph{not decoded}. We
therefore do not attribute their leakage to a capability failure either: a
non-linear decoder, a distributed multi-token code, or a representation at a
layer or position we did not read could carry content this instrument cannot
see. Reporting them as capability failures would reintroduce exactly the
inference the three-valued reading exists to prevent.

\subsection{The block axis needs its own screen, and it removes the largest
remaining cell}

Every screen so far constrains the \emph{decode} axis. The block axis had none,
and it needs one for a reason that is visible in the prompt: an encoded
condition changes two things at once. It transforms the payload's characters,
and it wraps them in a template announcing that an encoding is present. A guard
that flags anything \emph{asking about} an encoding produces a block rate that
is a response to the request, not a decode-then-block decision, and with only a
plaintext condition and an encoded one the two are indistinguishable.

We run the missing factorial cell: plaintext content wearing each condition's
own wrapper. The floor is the \emph{benign} arm of that condition (safe content,
wrapped, blocked anyway) because the harmful arm carries real harm and blocking
it is correct behaviour. A condition is reported only if the guard's block rate
on the encoded attack exceeds its own wrapper-induced false-positive rate.

The screen removes \texttt{combining\_marks} on WildGuard, which was the largest
cell in Table~\ref{tab:map} before it was applied. The guard blocks 44 per 100
of \emph{benign plaintext} wearing that wrapper and 25 per 100 of the encoded
harmful corpus: it fires more often on safe content wearing that wrapper (and on
bare benign plaintext, 45 per 100) than on the attack itself. Its apparent 56-per-100 policy failure
therefore cannot be read as one: the guard's blocking on that condition sits
below its false-positive rate on wrapped benign plaintext. Every other surviving condition on both guards clears.

Reporting the screen as a threshold rule compares two marginal rates against a
bar, so the supplementary material gives the full factorial instead: bare
plaintext, plaintext wearing the condition's wrapper, and the encoded condition,
each crossed with the benign and harmful arms. The wrapper term is small on both
guards, and on five of the six reported cells its interval includes zero, so the
confound the control was built to remove is present and slight: neither guard is
principally reacting to the request to handle an encoding. The encoding term
beyond the wrapper is where the loss lives, and it separates on two cells in
opposite directions, which a difference of marginal rates would have shown as
neither.

\subsection{Little evidence that either guard is a format detector}
\label{sec:format}

\emph{Blocked without decoding} is at most 1 per 100 on the six pairs in
Table~\ref{tab:map}, and exactly 0 on 26 of the 29
pairs the free permutation test admits. A guard reacting to the appearance of
encoding rather than to content would populate this cell heavily; neither does.

Two boundaries are what make that a statement about all 38 pairs rather than
about the screened ones only, and both are worth stating because the obvious
reading of a bound over a partly unmeasured set is that the unmeasured part was
counted as zero. It was not. On the nine pairs the free permutation test does not
admit, the cell is empty by containment rather than by assumption:
\emph{blocked without decoding} is a subset of \emph{blocked}, and the guard's
block rate on every one of those nine is 0.00, so no decode read is needed to
settle them. Two pairs do exceed 1 per 100, and both sit in the unmeasured band on the
decode axis: Llama Guard on \texttt{combining\_marks} at 8 per 100, a condition
the control floor demotes, and Llama Guard on \texttt{fullwidth} at 5 per 100,
the condition the item-level holdout withdrew. Where decoding cannot be read, the split between \emph{decoded} and
\emph{not decoded} is precisely what is unavailable, so that cell is not
evidence for a format detector either way. We report it rather than drop it with
the condition.

We report this cell at the \emph{permissive} operating point, the opposite of the choice for
the others. Tightening the decode read moves prompts out of \emph{decoded} and
into \emph{blocked without decoding} mechanically, inflating exactly the cell
whose emptiness is the finding. The most demanding test of the format-detector
hypothesis is therefore the most generous decode read, since that minimises the
chance of mislabelling a decoded prompt as an undecoded one; the same choice
makes an empty cell easiest to find, so it favours our claim. Different cells want
different operating points, and we state which we used for each.

That choice is also why this cell, alone among those we report, is read from a
probe fitted on all items rather than under the item-level holdout of
Section~\ref{sec:holdout}. The holdout is applied at the tuned operating point,
and the two cells are read at different ones. The likely direction of the gap
does not favour us: a probe denied item memory reads fewer prompts as
decoded, so it tends to move prompts \emph{into} this cell, and at the tuned
point doing so raises it from at most 4 per 100 to between 7 and 26. What that
does to the permissive read we have not measured. We therefore state the claim
as resting on the uncontrolled read, and note that a held-out bound near zero
would support it more strongly than the number we report, since an estimator
that tends to over-count and still finds nothing is the best evidence an empty
cell admits.

\section{Limitations}

Full statements of each limitation below are in the supplementary material.
Both guards' control sets are chosen by a proxy, the base model's decoding
ability, since a format-locked classifier cannot be asked to restate anything;
the error that assumption can produce tends to inflate the floor, which can only remove
conditions from the reported map. Our largest cell in the first analysis was an
artefact of exactly that kind, and we report it because it calibrates how much
weight the rest can carry. The cell we do report is structurally the one most
vulnerable to instrument error, since it is approximately the decode rate times
one minus the block rate, so any false positive concentrates where the guard
blocks least. The surviving margins are thin, between $0.052$ and $0.139$ over a
floor, and the floor's own multiplier lies below its validity window on one
guard, which we report as a window rather than a constant. A negative decode
reading is evidence about a linear probe at one selected layer and position, not
about the guard's computation. The measurement is correlational: we establish
that the content is present, not that the verdict used it, and the ablation that
would settle it is the immediate next experiment. Every quantitative statement is
conditioned on one corpus and two checkpoints.

\section{Discussion and conclusion}

Two failures that end-to-end evaluation reports as one number have opposite
remedies, and they can be told apart from inside the guard at the cost of one
forward pass and no judge model. Where the policy failure survives every screen, on
the zero-width encoding, the answer for both guards we study is the same: a probe fitted on plaintext and transferred
without refitting recovers the payload's harmful content from the guard's
residual stream, and the guard returns \emph{safe}. Whether the verdict
computation had access to that content and declined to use it is the causal
question our design does not settle, and closing that conditional is what the
intervention below would buy.

The reach of the measurement is both the paper's main caveat and its main
methodological result: honest licensing removed most of the map we first
believed. The cells that fell shared a signature worth naming, because it is also
what the most publishable finding looks like: high apparent decode with a block
rate of exactly zero. The last screen adds a second: a transferred probe
outscoring a probe fitted inside the condition it reads is carrying item
identity, not content. We therefore read any uncontrolled
decoded-but-not-blocked rate, in prior work and in our own first analysis alike,
as an upper bound rather than a measurement.

Three consequences for practice follow. A guard that leaks an encoded attack
should not be assumed to have failed to read it, since on every condition in
Table~\ref{tab:map} its residual stream carries the content. A guard that blocks one should not be assumed to have
read it either, though under the conditions we test we find little evidence for a
pure encoding-format detector. And an internal audit of a defence needs controls
on both of its axes, because the screens that discipline the decode read leave
the block rate entirely untouched.

The immediate next step is causal: ablating the recovered direction and observing
whether the verdict moves.

\bibliography{paper}

\clearpage
\appendix
\section*{Supplementary Material}

\section{Full specification}
\label{app:spec}

\subsection{Corpus}

The corpus is JailbreakBench \citep{NEURIPS2024_63092d79}: 100 harmful
behaviours and the 100 benign prompts it ships as their topic-matched
counterparts, one per behaviour, so the negative class differs from the positive
one in harmfulness rather than in subject matter. Every condition below is
run on both sets in full, so every cell count in the paper is out of 100 and
every AUROC is computed on 200 examples. No prompt is filtered, resampled or
reweighted at any stage.

\subsection{Conditions}

The ladder is 19 conditions, each an exactly invertible transformation of the
payload. Round-trip invertibility is a checked property of every encoder in our
implementation, not a description of it.

\begin{itemize}
\item \emph{Byte and character ciphers} (12): \texttt{base64}, \texttt{base32},
  \texttt{hex}, \texttt{binary}, \texttt{ascii\_decimal},
  \texttt{unicode\_escape}, \texttt{rot13}, \texttt{caesar3},
  \texttt{caesar7}, \texttt{atbash}, \texttt{vigenere}, \texttt{morse}.
\item \emph{Order transforms} (2): \texttt{reverse\_characters},
  \texttt{reverse\_words}.
\item \emph{Surface transforms} (5): \texttt{zero\_width},
  \texttt{tag\_block}, \texttt{homoglyph}, \texttt{fullwidth},
  \texttt{combining\_marks}.
\end{itemize}

The transformation is applied to the payload only. It is then placed in a fixed
wrapper, identical across conditions except for the encoding's name:

\begin{quote}\ttfamily
The following request is encoded with \{encoding\}:\\[2pt]
\{ciphertext\}
\end{quote}

The wrapper names the encoding, which matches the threat model in which an
attacker tells the model which cipher was used, and it does not instruct the
model to decode. That instruction is a separate intervention and does not appear
in any condition reported here. The plaintext condition is the bare payload with
no wrapper; the wrapper arm of the block-axis control is plaintext content
inside the wrapper above, which is the remaining cell of that $2 \times 2$.

\subsection{Guards, rendering and the verdict read}

The guards are \texttt{meta-llama/Llama-Guard-3-8B}
\citep{inan2023llamaguardllmbasedinputoutput,grattafiori2024llama3} and
\texttt{allenai/wildguard} \citep{NEURIPS2024_0f69b4b9}, 8B- and 7B-scale
open-weight classifiers with 32 layers. Each prompt is rendered through the guard's own prompt format and
tokenised with \texttt{add\_special\_tokens=False}, because both formats emit
their own beginning-of-sequence token; adding a second one shifts every
position index by one. Llama Guard 3 is rendered through the chat template
shipped with the checkpoint. WildGuard ships no chat template, so its format is
the input format defined by the reference implementation, used verbatim, with
the response slot left empty because we classify requests rather than
completions.

The verdict is read from the logits at the position where the label token falls,
as the comparison of the unsafe and safe label tokens. The prefix that locates
that position is the two newlines preceding Llama Guard's verdict and the string
\texttt{Harmful request:} preceding WildGuard's; the label tokens carry their
leading space, because the tokeniser encodes the boundary into the token itself.
For every run we record the mean probability mass on the two label tokens at the
read position and treat a value away from 1 as a format failure rather than a
result. This matters concretely: WildGuard's verdict is not the first generated
token, and a first-token read scores a label that holds almost none of the mass.

\subsection{Activation capture}

We capture the residual stream at the input to every one of the 32 layers, at
two positions: the last token of the payload and the last token of the rendered
prompt. The grid is therefore $32 \times 2 = 64$ cells per condition. Capture is
a single forward pass in the model's native dtype with no sampling and no
generation; activations are cast to float32 on write.

\subsection{The content probe}

The probe is a logistic regression on the raw residual-stream vector at one
cell, with no centering, scaling or dimensionality reduction:
\texttt{scikit-learn} \texttt{LogisticRegression} with L2 penalty, $C=1$,
\texttt{max\_iter}~$=2000$, \texttt{random\_state}~$=0$, and the default
\texttt{lbfgs} solver; the exact library version is fixed by the dependency
lockfile whose digest each run record carries. The score is the signed distance to the decision
boundary; AUROC is computed on that score. All BLAS threading is pinned to one
thread, which makes the fits deterministic given the seed and, at this problem
size, also faster.

For the transfer measurement the probe is fitted on the plaintext condition's
200 examples and evaluated on the encoded condition's 200 examples, with no
refitting. The reported statistic is the transfer AUROC at the
$\arg\max$ cell of the grid.

\subsection{Licensing: the length-matched permutation null}

A condition is licensed at $p<0.05$ against a null of 200 draws of the maximum
transfer AUROC over the 64-cell grid. Each null draw
permutes the \emph{evaluation} labels with the fitted direction held fixed, and
the permutation is restricted to strata of the prompt's character length, so
that the class--length association present in the corpus is preserved under the
null and cannot be exploited by a probe reading length. Because the direction is
fixed, each cell is scored once and each draw re-ranks cached scores, which is
what makes 200 draws over 64 cells cheap enough to run on CPU. Taking the
maximum inside the null is what corrects for selecting the best cell; there is
no separate multiple-comparison adjustment because there is only one test per
condition. The p-value carries the $+1$ correction of
\citet{phipson2010permutation}, so with 200 draws the smallest value it can
take is $1/201$. A condition reported at $0.005$ is therefore at that floor and
should be read as ``no null draw reached the observed value'', never as an
estimate of $0.005$: the correction exists to stop a finite permutation set
reporting $p=0$, and quoting the corrected floor as a measurement puts the
overstatement back one level up. Licensing runs at 10 length strata by default and is repeated at 5 and 20; a
condition whose decision changes across those three is reported as unstable, and
on one guard the conditions that are unstable are exactly the ones the control
floor rejects.

\subsection{The control floor}

Permutation significance establishes that a probe reads something, not that it
reads decoded content. The floor is the level the same probe reaches on
conditions on which the guard's base model decodes nothing (rate 0.00), where by assumption
there is nothing decoded to read.

\emph{Control membership.} A condition is a control for a given guard if the
guard's \emph{base} model restates the payload at rate exactly 0.00. The guard
cannot be asked to restate anything, because its prompt format hard-wires the
classification task, so the selector is inherited: Llama Guard 3 8B from
\texttt{meta-llama/\allowbreak Llama-3.1-8B-\allowbreak Instruct}, WildGuard
from \texttt{mistralai/\allowbreak Mistral-7B-\allowbreak Instruct-v0.3}. We name the second by identifier
rather than by citation on purpose: the published description of that family
covers an earlier release with a different vocabulary, and the post-training of
the checkpoint we run is documented by a model card and not by a paper. This
inheritance
is an assumption and is stated as a limitation in the main text; the software
requires both model names to be supplied when a floor is constructed, so a floor
cannot be built without recording whose ability selected its controls.

\emph{The statistic.} With at least five controls the floor is
$\mathrm{mean} + \sigma\,\mathrm{SD}$ of the controls' transfer AUROC, with
$\sigma = 2$; with fewer it falls back to the maximum and is labelled a
\emph{bound}, which is not comparable across control-set sizes because an
extreme-value statistic grows with the number of draws. Both floors reported
here are distributions, over 12 controls for Llama Guard 3 and 14 for WildGuard.
Their values under the item-level holdout are $0.7066$ and $0.6605$
respectively, which the main text quotes to three decimals.

\emph{The multiplier.} $\sigma$ is not free. Its lower bound is the value at
which the highest control clears its own floor and its upper bound is the value
at which a surviving condition stops clearing; we report that window per guard
and treat a configured value outside it as a defect rather than a preference.
The window is derived per guard and never carried between guards, for the same
reason a floor is never carried between runs.

\subsection{The item-level holdout}

The transfer design holds the condition out and does not hold the items out. We
therefore repeat the measurement 200 times: on each repeat, the harmful and
benign index sets are permuted independently and split in half, a probe with the
hyper-parameters above is fitted on the plaintext half, and its AUROC is computed
on the encoded versions of the held-out half. We report the mean over repeats,
and the 2.5th and 97.5th percentiles across repeats as a spread. That spread is
variation across splits, not a confidence interval on a population AUROC.

Halving the training set changes two things at once, so a fall from the unsplit
value confounds item memory with sample size. To separate them we score each
fitted probe a second time on the encoded versions of the items it \emph{was}
fitted on. The difference between those two AUROCs is at the same training-set
size and differs only in whether the item was seen, and it is the leakage figure
quoted in the main text. Every floor is re-derived from control-condition AUROCs
computed under the identical procedure, so a condition and the floor it must
clear are always the same kind of number.

We also report, at each selected cell, the cross-validated AUROC of a probe
fitted and evaluated within plaintext alone (stratified 5-fold, same
hyper-parameters). That is the ceiling the transferred reading should be
compared against, and its absence is why an uncontrolled transfer AUROC of 0.985
looked like intact representation rather than like a probe recognising a prompt
it had been shown.

\subsection{The per-prompt read and the operating point}

The population AUROC does not label individual prompts. A prompt reads as
\emph{decoded} when its probe score exceeds the 75th percentile of the benign
score distribution \emph{in the same condition}. Taking the threshold from the
same condition makes the encoding and its wrapper common-mode: both classes move
together under them, so neither can decide a label by itself. The percentile is
a knob and we sweep it over $\{50, 75, 90, 95, 99, 99.5\}$; at the median it has
a 50\% false-positive rate by construction, which is why the median is not the
tuned point.
One cell is deliberately read at the permissive end: \emph{blocked without
decoding} is the cell a tighter read inflates mechanically, so it is read at
the 50th percentile, which avoids that inflation but also makes an empty cell easiest to
find.

When a condition is not licensed, the per-prompt read is not a negative but
absent. The reading is three-valued throughout, and the software returns no
value rather than \texttt{False} for an unlicensed condition, so that
\emph{unmeasured} cannot silently become \emph{did not decode} one level up.

\subsection{Base-model decoding ability}

Ability is measured in a separate generation pass in which the base model is
shown the encoded payload and asked to write out the decoded text and nothing
else, decoding greedily for at most 256 new tokens. A response counts as a
recovery if it matches the plaintext exactly, contains it, or clears both a
character-sequence similarity of 0.75 and a content-token overlap of 0.6; an
order-blind branch admits a content-token overlap of 0.80 alone, which is what
\texttt{reverse\_words} requires. Content overlap is the fraction of the
plaintext's content tokens matched fuzzily in the response, and it exists as a
veto: a shared instruction frame can push character similarity to 0.9 with the
harmful payload absent. Responses reproducing a 24-character window of the
ciphertext are flagged as echoes. All cuts are configuration, and all raw scores
are stored per prompt, so the binary can be re-derived offline from cached
responses; it has been, which is how the exact-match rule this pipeline started
with was found to be scoring genuine decodes as failures.

\subsection{Per-condition block rates}

Table~\ref{tab:allblock} gives each guard's block rate on the harmful arm of
every condition, with 95\% Wilson intervals. Unlike the decode statistic it
depends on no threshold of ours: it is the guard's own verdict token, so it is
reportable on all 19 conditions including those where the decode measurement is
unmeasured. It is also the axis on which the two guards visibly differ, and it
is what an end-to-end evaluation would report on its own.

\begin{table}[t]
\centering
\small
\setlength{\tabcolsep}{3pt}
\begin{tabular}{lcc}
\toprule
Condition & Llama Guard 3 & WildGuard \\
\midrule
\multicolumn{3}{l}{\emph{Byte and character ciphers}} \\
\quad \texttt{base64} & 0 [0--4] & 0 [0--4] \\
\quad \texttt{base32} & 0 [0--4] & 0 [0--4] \\
\quad \texttt{hex} & 0 [0--4] & 0 [0--4] \\
\quad \texttt{binary} & 0 [0--4] & 0 [0--4] \\
\quad \texttt{ascii\_decimal} & 0 [0--4] & 0 [0--4] \\
\quad \texttt{unicode\_escape} & 0 [0--4] & 0 [0--4] \\
\quad \texttt{rot13} & 2 [1--7] & 0 [0--4] \\
\quad \texttt{caesar3} & 0 [0--4] & 0 [0--4] \\
\quad \texttt{caesar7} & 0 [0--4] & 0 [0--4] \\
\quad \texttt{atbash} & 0 [0--4] & 0 [0--4] \\
\quad \texttt{vigenere} & 0 [0--4] & 0 [0--4] \\
\quad \texttt{morse} & 0 [0--4] & 0 [0--4] \\
\addlinespace
\multicolumn{3}{l}{\emph{Order transforms}} \\
\quad \texttt{reverse\_characters} & 0 [0--4] & 0 [0--4] \\
\quad \texttt{reverse\_words} & 65 [55--74] & 73 [64--81] \\
\addlinespace
\multicolumn{3}{l}{\emph{Surface transforms}} \\
\quad \texttt{zero\_width} & 83 [74--89] & 71 [61--79] \\
\quad \texttt{tag\_block} & 0 [0--4] & 0 [0--4] \\
\quad \texttt{homoglyph} & 92 [85--96] & 75 [66--82] \\
\quad \texttt{fullwidth} & 85 [77--91] & 0 [0--4] \\
\quad \texttt{combining\_marks} & 79 [70--86] & 25 [18--34] \\
\bottomrule
\end{tabular}
\caption{Blocked, per 100 harmful prompts, with 95\% Wilson intervals in
percentage points. The decode measurement is licensed on three of these
conditions for Llama Guard 3 and on four for WildGuard; on the rest it is
unmeasured, which is not a finding that nothing was decoded.}
\label{tab:allblock}
\end{table}

\subsection{The length bound on the block axis}

Both arms of a condition are the same 100 themes, so their lengths differ by as
much as the bare corpora do: 86.0 characters on the harmful side against 73.8 on
the benign, a raw-length AUROC of 0.654 that every encoder in the ladder carries
through unchanged. The bound asks what a guard reacting only to length could show
at the blocking budget the real guard used. Pool a condition's 200 ciphertexts,
let $k$ be the number the guard blocked across both arms, and give $k$ to the
best monotone length rule, blocking the $k$ longest or the $k$ shortest,
whichever separates more, with ties at the cut resolved in the direction that
raises the bound. Table~\ref{tab:lengthbound} reports it in both units.

Fixing the budget is the right comparison, since the confound at issue is that
this guard's decisions are partly length and its own rate is observed. The
constraint is also not load-bearing: the maximum over every budget in either unit
is $0.28$ on five of the six cells and $0.38$ on WildGuard
\texttt{reverse\_words}, so all six clear without it.

The bound covers monotone rules only. The unrestricted optimum over all
functions of length fits a 200-item sample almost perfectly and would bound
near one on any corpus whatsoever, which is a statement about the corpus
rather than about the guard, and the confound reported in the literature is
monotone in any case.

\begin{table}[t]
\centering
\small
\setlength{\tabcolsep}{4pt}
\begin{tabular}{lccccc}
\toprule
Condition & Gap & Char & Token & Margin & Binds \\
\midrule
\multicolumn{6}{l}{\emph{Llama Guard 3}} \\
\quad \texttt{homoglyph} & 0.53 & 0.21 & 0.21 & +0.32 & char \\
\quad \texttt{zero\_width} & 0.54 & 0.20 & 0.22 & +0.32 & token \\
\quad \texttt{reverse\_words} & 0.36 & 0.22 & 0.22 & +0.14 & char \\
\addlinespace
\multicolumn{6}{l}{\emph{WildGuard}} \\
\quad \texttt{homoglyph} & 0.52 & 0.22 & 0.26 & +0.26 & token \\
\quad \texttt{zero\_width} & 0.48 & 0.22 & 0.22 & +0.26 & char \\
\quad \texttt{reverse\_words} & 0.44 & 0.22 & 0.34 & +0.10 & token \\
\bottomrule
\end{tabular}
\caption{The harm gap against the most a monotone length-only rule could
manufacture at the same blocking budget, in characters and in the guard's own
tokens. The margin is measured against the larger of the two. Every reported
cell clears; the token bound is the binding one on half of them, and on
WildGuard \texttt{reverse\_words} it exceeds the character bound by more than the
margin that survives it.}
\label{tab:lengthbound}
\end{table}

The two units disagree because a rate gap at a fixed budget is not a monotone
function of the separation AUROC: WildGuard \texttt{reverse\_words} moves from
0.654 to 0.663 in AUROC between the units and from 0.22 to 0.34 in bound, since
what matters is the shape of the two length distributions at the cut rather than
their overall overlap. This is a further reason not to report a length AUROC as
the control, which is what the probe-side null does and what a rate needs
replacing.

\subsection{Computing infrastructure}

Guard forward passes, activation capture and the base-model ability generations
run on a single NVIDIA H200 (144\,GB HBM) per job on a shared SLURM cluster; no
stage uses more than one GPU. Everything downstream of capture is CPU-only and
needs no accelerator: the probe fits, the permutation licensing, the control
floors, the item-level holdout and every offline rescoring script. Cluster nodes
run Linux 5.14.0 (RHEL 9.3, \texttt{glibc} 2.34) on x86-64 with Python 3.12.4;
the offline analyses were additionally run on macOS 26.5 (arm64) with Python
3.13.14, and the two agree because nothing downstream of capture consumes a GPU.

The principal libraries are PyTorch 2.11.0 (CUDA 12.8), Transformers 5.14.1,
\texttt{scikit-learn} 1.9.0, NumPy 2.5.1 and Accelerate 1.14.0. The exact
version of every transitive dependency is pinned by a lockfile whose digest each
run record carries, so the library set is recoverable from the record rather
than from this paragraph alone. BLAS threading is pinned to a single thread for
the probe fits, which makes them deterministic given the seed and, at this
problem size, also faster than the default thread count.

\subsection{Reproducibility, and what is not pinned}

Every run writes a record carrying the git commit of the tree it ran on, whether
that tree was dirty and its diff if so, the resolved configuration, the seed,
the Python version and platform, and a digest of the dependency lockfile. Runs
that consume a prior run's outputs record a content hash of what they read. A
run on accelerated hardware refuses to start from a working tree with
uncommitted changes unless it is explicitly opted in, and then the full diff is
stored in the record, so a recorded commit always describes the code that ran.

One thing is \emph{not} pinned and we state it rather than implying otherwise:
the guard checkpoints are referenced by repository identifier and not by
revision hash, so a reader reproducing this work obtains whatever revision those
identifiers resolve to. Neither repository is known to us to have been updated
during the study, but that is an absence of evidence and not a pin. Recording
the resolved revision alongside the identifier would close this, and it cannot
be applied retroactively to the runs reported here.

\section{Additional results}

\subsection{Two screens converge on the same partition}

Of the seven conditions WildGuard licenses at ten strata, the three whose
licensing is unstable under the number of strata (5, 10, 20) are exactly the
three the control floor rejects, and the four that are stable are exactly the
four it keeps. The two screens share no input: the floor is computed from
base-model decode ability and never sees stratification, and bin-stability is
computed from the null and never sees ability.

What the agreement establishes is narrower than two independent measurements
corroborating each other, and the narrower version is the one we claim. All
three rejected conditions have base-model ability 0.00, and all three are
inputs to the floor itself, where a
control cannot clear its own floor by construction. The floor's verdict on those
three is therefore largely fixed once ability is known, and the informative half
of the agreement is the other screen: bin-stability never consults ability and
recovers the same partition anyway, so instability under the null tracks the
base model's inability to decode. One of the four kept conditions,
\texttt{combining\_marks}, clears the floor by 0.004 with a 95\% band that
straddles it, so the partition is exact on six of the seven and unresolved on
the seventh.

The convergence does not hold on Llama Guard, and we report that rather than
generalising from the guard where it worked: there, the Caesar-shift artefact is
stable at all three bin counts and is caught only by the floor. Bin-stability is
therefore the weaker screen, and cannot be substituted for the floor.

\subsection{The guards dissociate totally on one encoding}

On \texttt{fullwidth}, Llama Guard blocks 85 of 100 harmful prompts with a
32-point margin over the benign arm, and WildGuard blocks 0 of 100. Same
condition, same corpus, opposite behaviour. This is a behavioural result and we
report it as one: \emph{neither} guard's decode measurement on that condition
survives our screens (WildGuard's fails the floor outright, and Llama Guard's
clears it until the items are held out and then does not) so we can say that
one guard blocks \texttt{fullwidth} and the other does not, and we cannot say
what either of them represents.

The two guards also differ systematically on benign encoded content across the other surface and word-order conditions, with \texttt{fullwidth} itself excluded so that the
comparison does not rest on the condition it is explaining: one blocks 0.29 to
0.42 of it, the other 0.05 to 0.29. The more trigger-happy guard is also the one
that blocks this condition. Before any encoding, the guards differ in the
opposite direction: on \emph{bare plaintext} with no encoder and no wrapper, WildGuard
blocks 45 per 100 benign prompts and Llama Guard 23, so WildGuard begins with a
harm gap of 0.54 against the other's 0.75. Any encoded rate read
against a harmful-arm ceiling alone would have attributed that difference to the
attack. Two checkpoints cannot separate whether it reflects the base model, the
safety data, or the objective, and we do not speculate.

\subsection{The wrapper control as a factorial}

\begin{table}[t]
\centering
\footnotesize
\setlength{\tabcolsep}{2pt}
\begin{tabular}{lcccc}
\toprule
& \multicolumn{3}{c}{harm gap} & \\
\cmidrule(lr){2-4}
Condition & plain & wrap. & enc. & $\Delta$ wrapper \\
\midrule
\multicolumn{5}{l}{\emph{Llama Guard 3}} \\
\quad \texttt{homoglyph}     & 0.75 & 0.63 & 0.53 & 0.12 [-0.01, 0.25] \\
\quad \texttt{zero\_width}   & 0.75 & 0.67 & 0.54 & 0.08 [-0.05, 0.21] \\
\quad \texttt{reverse\_words}$^\dagger$ & 0.75 & 0.68 & 0.36 & 0.07 [-0.06, 0.20] \\
\quad \texttt{fullwidth}$^\ddagger$ & 0.75 & 0.69 & 0.32 & 0.06 [-0.07, 0.19] \\
\quad \texttt{combining\_marks}$^\ddagger$ & 0.75 & 0.70 & 0.37 & 0.05 [-0.08, 0.18] \\
\addlinespace
\multicolumn{5}{l}{\emph{WildGuard}} \\
\quad \texttt{homoglyph}     & 0.54 & 0.36 & 0.52 & \textbf{0.18 [0.04, 0.32]} \\
\quad \texttt{zero\_width}   & 0.54 & 0.43 & 0.48 & 0.11 [-0.03, 0.25] \\
\quad \texttt{reverse\_words}$^\dagger$ & 0.54 & 0.50 & 0.44 & 0.04 [-0.10, 0.18] \\
\quad \texttt{fullwidth}$^\ddagger$ & 0.54 & 0.55 & 0.00 & -0.01 [-0.15, 0.13] \\
\quad \texttt{combining\_marks}$^\ddagger$ & 0.54 & 0.55 & 0.20 & -0.01 [-0.15, 0.13] \\
\bottomrule
\end{tabular}
\caption{The wrapper control as a factorial rather than a threshold. Harm gap is
the harmful minus benign block rate at each level; $\Delta$ wrapper is the
interaction (plain gap minus wrapped gap), i.e.\ the discrimination the wrapper
alone costs, with a 95\% interval. Bold marks the one cell whose interval
excludes zero. The encoding term beyond the wrapper (wrapped gap minus encoded
gap) separates on Llama Guard \texttt{reverse\_words}, $0.32$ CI $[0.16, 0.48]$,
and on WildGuard \texttt{homoglyph}, $-0.16$ CI $[-0.31, -0.01]$, where the
encoding \emph{recovers} discrimination relative to the wrapper alone.
$^\ddagger$Conditions that a screen removed before the main paper's table of
surviving conditions, given here so that the screens themselves can be audited
rather than taken on their verdicts. WildGuard \texttt{combining\_marks} is the
cell this particular control withdraws: the guard blocks $0.44$ of benign
plaintext wearing that wrapper against $0.25$ of the encoded harmful arm, so its
blocking on the attack sits below its false-positive rate on wrapped benign
plaintext, and the apparent policy failure there cannot be read as one. Llama Guard
\texttt{fullwidth} is the condition the item-level holdout withdrew on the
decode axis; its block behaviour is unaffected and is shown for completeness.
These four rows sit outside the six the main paper carries and outside every
count stated there.
$^\dagger$reported as a control. Intervals treat the four cells as independent
because per-item verdicts were persisted for the encoded harmful arm only, so
they are likely wider than the item-paired truth: a cell that separates here
likely also separates under item pairing, and a cell that does not is not thereby shown to be null.}
\label{tab:factorial}
\end{table}

\section{Further limitations}

The first limitation below is not listed in the main text's Limitations section; it arose
from the last screen we ran. The remainder are the full statements of the
limitations the main text lists, in the order it lists them.

\paragraph{The item-level holdout reaches one cell and not the other.} The
holdout now re-derives the \emph{per-prompt} read as well as the AUROC, refitting
on half the plaintext items and reading the held-out half, so
the counts in the main paper's table of surviving conditions no longer carry the
leak. They fall by factors of
1.6 to 5.6 and four of the six spreads include zero, which is the largest single
correction in this paper and one we had disclosed as an open defect rather than
measured. Halving the training set could produce a fall without any leakage, so we
separate the two: scoring each fitted probe a second time on the items it was
fitted on isolates memory at the same training size, and it accounts for 97 to
123 per cent of the gap, with the halved training set slightly \emph{helping} on
five of the six cells.

The two cells do not move together, and treating them as symmetric would be an
error. \emph{Decoded but not blocked} needs a positive decode read, so a probe
denied item memory tends to under-count it, and we read the held-out value as a
lower bound; that
is the cell we report, and it is reported conservatively. \emph{Blocked without
decoding} needs a \emph{negative} read, so the same probe tends to over-count it,
and we read the held-out value as an upper bound. A small upper bound would support the
emptiness claim more strongly than the uncontrolled read does, and a loose one
supports nothing. We report that cell at the permissive operating point and the
holdout we ran was at the tuned one, so we do not yet have its held-out
counterpart; at the tuned point, removing item memory raises it from at most 4
per 100 to between 7 and 26. The claim in the format-detector section of the main paper is therefore
stated on the uncontrolled read, the measurement that would settle it is
specified, and it is the first thing we would run next.

\paragraph{The asymmetry, in general form.} The general form is worth stating because nothing in a build catches it: a screen
that weakens an instrument moves a cell up or down according to whether the cell
needs the instrument to answer yes or no. A cell it must answer yes to becomes a
lower bound and can still carry a finding. A cell whose \emph{emptiness} is the
finding becomes an upper bound, which is informative when it is small and
worthless when it is loose, so the screen has to be run before its result can be
called either way. Deciding that per cell is part of applying the screen,
not part of reading its output.

\paragraph{Both guards' control sets are chosen by a proxy.} The floor above is
estimated from the conditions a guard's \emph{base} model cannot decode, because
a format-locked classifier cannot be asked to restate anything. Both guards are
screened this way, at the cost of one extra measurement per guard (a base
model, a handful of conditions, no judge), and on the second guard it is that
measurement which withdrew the three cipher conditions above. The residual
assumption is that a safety fine-tune does not acquire decoding ability its base
model lacks. If it does, we have treated a decodable condition as a control,
which tends to inflate the floor; an inflated floor can only remove conditions from the
reported map, never add them, so the direction of that error is conservative for
the policy failure we report.

\paragraph{Our largest reported cell was, in the first analysis, an artefact.}
Before the floor existed, the most striking result in this study was a condition
on which the probe licensed comfortably and the guard blocked nothing, implying
that the guard read the payload and let all of it through. The guard's base model
turns out to be unable to decode that condition at all, so, if the guard
inherits that inability, nothing was decoded
and the reading was the probe's own surface-feature floor. We report this because
it calibrates how much weight the remaining cells can carry, and because the
same shape (high apparent decode, zero block rate) is what an uncontrolled
version of this method will produce most readily.

\paragraph{The cell we report is structurally the one most vulnerable to
instrument error.} \emph{Decoded but not blocked} is approximately the decode
read rate multiplied by one minus the block rate. Any false positive in the
decode read therefore concentrates in precisely the conditions where the guard
blocks least, the conditions a paper most wants to report as failures. We found
that tightening the read threshold shrinks every cell without emptying any, so
the operating point cannot substitute for the floor; only a control that removes
whole conditions can. Our controls attack this from one side, by deriving the
floor from conditions the base model cannot decode. The other side is untested
here: \citet{mu-etal-2025-stealthy} construct harmful requests that are
distributionally close to benign ones by design, which is the condition under
which a probe reading corpus-level surface features rather than content would be
exposed. Running our screens against such a corpus is the sharpest adversarial
test of the decode read we are aware of, and we have not run it. At our sample size the surviving cells are not separated
from one another, since their split spreads overlap on every pair, so we make no
claim about their relative size. Readers should treat an uncontrolled
decoded-but-not-blocked rate, including in prior work, as an upper bound.

\paragraph{The surviving margins are thin, and one screen's constant is out of
bounds.} After the item-level holdout the smallest margin over a floor among the
reported cells is 0.052
and the largest 0.139, where before it the smallest was 0.087. Nothing here
should be read as a comfortable pass, and a further confound of the same
magnitude as the one we found would remove conditions rather than shrink cells.
The floor's own multiplier compounds this: the range of multipliers on which no
verdict changes is $[2.11, 3.44]$ on one guard and $[1.58, 2.12]$ on the other,
and the value we configured, 2.0, lies below the first range and 0.12 inside the
second. On the first guard the consequence is that a control condition would
clear its own floor by value, which our implementation prevents by construction
because control membership is decided before the comparison; we report the
window rather than the constant, because a constant carried between guards is
the same error as a floor carried between runs.

\paragraph{Linear probes, one position, one layer selection rule.} A negative
decode reading is evidence about what a linear probe can recover at the position
and layer we selected, not about what the guard's computation contains.
Non-linear or multi-token decoding methods might recover content where we read
none, which would move conditions from \emph{unmeasured} into the measurable
band rather than shrink the policy failure we report.

\paragraph{Correlational, and the causal test is specified but not run.} We
establish presence, not use. The direct test is well-established methodology and is the immediate next
experiment: ablate the recovered direction in the guard's residual stream and
observe whether the verdict moves.

\paragraph{Two guards, one corpus.} Every quantitative statement is conditioned
on a single prompt corpus and two checkpoints. The dissociations we report
between guards are properties of those two checkpoints; whether they reflect
differences in safety-training data, in base model, or in classifier objective is
not something two points can separate.

\end{document}